\documentclass[aps,prb,floatfix,reprint,showpacs,superscriptaddress,longbibliography]{revtex4-2}

\usepackage{mathtools,amssymb,graphicx}
\usepackage[plainpages=false,pdfpagelabels,colorlinks=true,linkcolor=red,urlcolor=blue,citecolor=blue,pdftitle={Title},pdfauthor={},pdfdisplaydoctitle=true,pdfduplex=DuplexFlipLongEdge]{hyperref}
\usepackage{soul} 
\usepackage[dvipsnames,usenames]{color}
\usepackage[normalem]{ulem}
\usepackage{graphicx}
\usepackage{booktabs}
\usepackage{xcolor}
\usepackage{bbold, verbatim} 
\usepackage{float}

\emergencystretch=\maxdimen
\usepackage[colorinlistoftodos,prependcaption,textsize=tiny]{todonotes}
\usepackage{cprotect}
\usepackage{siunitx}
\graphicspath{ {./images/} }
\newcommand{\ie}{i.\,e.}

\newcommand{\eg}{e.\,g.}

\begin{document}
\title{Hydrogen liquid-liquid transition from first principles and machine learning}

\author{Giacomo Tenti} 
\email{gtenti@sissa.it}
\affiliation{International School for Advanced Studies (SISSA),
Via Bonomea 265, 34136 Trieste, Italy}

\author{Bastian Jäckl} 
\affiliation{Department of Computer and Information Science, University of Konstanz, Box 188, 78457 Konstanz, Germany}

\author{Kousuke Nakano} 
\affiliation{Center for Basic Research on Materials, National Institute for Materials Science (NIMS), 1-2-1 Sengen, Tsukuba, Ibaraki 305-0047, Japan}
\affiliation{Center for Emergent Matter Science (CEMS), RIKEN, 2-1 Hirosawa, Wako, Saitama, 351-0198, Japan.}

\author{Matthias Rupp} 
\affiliation{Department of Computer and Information Science, University of Konstanz, Box 188, 78457 Konstanz, Germany}
\affiliation{Scientific Instrumentation and Process Technology unit, Luxembourg Institute of Science and Technology (LIST), Maison de l'Innovation, 5 avenue des Hauts-Fourneaux, L-4362 Esch-sur-Alzette, Luxembourg}

\author{Michele Casula}
\email{michele.casula@impmc.upmc.fr}
\affiliation{Institut de Minéralogie, de Physique des Matériaux et de Cosmochimie (IMPMC), Sorbonne Université, CNRS UMR 7590, MNHN, 4 Place Jussieu, 75252 Paris, France}

\begin{abstract}
The molecular-to-atomic liquid-liquid transition (LLT) in high-pressure hydrogen is a fundamental topic touching domains from planetary science to materials modeling.
Yet, the nature of the LLT is still under debate. To resolve it, numerical simulations must cover length and time scales spanning several orders of magnitude. 
We overcome these size and time limitations by constructing a fast and accurate machine-learning interatomic potential (MLIP) built on the MACE neural network architecture.  
The MLIP is trained on Perdew-Burke-Ernzerhof (PBE) density functional calculations and uses a modified loss function correcting for an energy bias in the molecular phase.
Classical and path-integral molecular dynamics driven by this MLIP show that the LLT is always supercritical above the melting temperature. 
The position of the corresponding Widom line agrees with previous \emph{ab initio} PBE calculations, which in contrast predicted a first-order LLT.
According to our calculations, the crossover line becomes a first-order transition only 
inside the molecular crystal region.
These results call for a reconsideration of the LLT picture previously drawn. 
\end{abstract}
\date{\today}
\maketitle

\paragraph*{Introduction.}
Pristine hydrogen is one of the most widely studied materials, both theoretically and experimentally~\cite{Bonitz2024}.
Indeed, despite being made of the simplest atomic constituent, it exhibits a surprisingly rich phase diagram.
The molecular-to-atomic transition happening upon liquid hydrogen compression is crucial for planetary science, in particular for understanding the interior of giant gas planets~\cite{Helled2020} and their magnetic fields~\cite{Connerney2017}.
This liquid-liquid transition (LLT) has been extensively investigated both experimentally~\cite{Weir1996,Fortov2007,Dzyabura2013,Ohta2015,Knudson2015,McWilliams2016,Zaghoo2016,Zaghoo2017,Zaghoo2018, Celliers2018, Jiang2020} and via numerical simulations~\cite{Scandolo2003, Bonev2004,  Delaney2006, Vorberger2007, Holst2008, Attaccalite2008, Lorenzen2010, Morales2010, Tamblyn2010, Mazzola2014, Yang2015, Pierleoni2016, Norman2017, Mazzola2018, Tian2019, Geng2019, Rillo2019, Hinz2020, Bryk2020, Cheng2020, Karasiev2021, Bergermann2024,Gliaudelis2025}.

As is often the case for high-pressure hydrogen, experiments based on static~\cite{Goncharov1995, Bassett2009} and dynamic~\cite{Nellis2006} compression give contrasting results, the dynamic 
experiments predicting a larger transition pressure. 
Uncertainty is also present in the numerical simulations. Results obtained with \emph{ab initio} molecular dynamics (AIMD) using density functional theory (DFT) show large variability with respect to the choice of the exchange-correlation functional. For instance, the transition pressure can vary by $200$~GPa when including long-range van der Waals corrections~\cite{Morales2014}. 

The nature of the LLT has been debated, with many first-principles simulations~\cite{Scandolo2003,Morales2010,Lorenzen2010,Pierleoni2016,Norman2017,Geng2019,Karasiev2021} suggesting a first-order LLT below a critical temperature~$T_c$ and pressure~$p_c$, based on the observation of kinks in the equation of state (EOS).
Given the long correlations in both time and space expected near the transition, the outcome of first-principles molecular dynamics (MD) simulations using DFT or quantum Monte Carlo (QMC) methods has been questioned because of the short time scales and/or the small system sizes considered. 
This could be reflected in the large variability of the predicted $T_c$ and $p_c$, which range from $T_c \sim 4000$\,K and $p_c \sim 30$\,GPa~\cite{Norman2017,Tian2019}, to $T_c \sim 1250$\,K and $p_c \sim 150 $\,GPa for the most recent simulations~\cite{Geng2019}.

In the past two decades, large-scale simulations have been made possible by the introduction of machine learning interatomic potentials~\cite{Behler2016} (MLIPs). 
These provide results at much lower computational cost, once trained on datasets generated with \emph{ab initio} methods. 
However, the derivation of an MLIP is a delicate step \emph{per se}, possibly introducing a residual bias. 
A recent MLIP model, based on the NequIP~\cite{bmsk2022q} neural network (NN) and trained on DFT data, gave a $T_c$ value smaller than previous estimates~\cite{istas2024liquid}. This model has a pressure bias of about 5 GPa in the EOS, once compared with the corresponding \emph{ab initio} predictions. Recent simulations based on another MLIP trained on the same DFT functional~\cite{Cheng2020} suggested that the LLT is instead a smooth crossover, even though the accuracy of the model has been criticized~\cite{Karasiev2021, Cheng2021}. 

In this work, we present results for the LLT obtained with MACE~\cite{Batatia2022}, a framework combining an equivariant message-passing NN with high-body-order messages, which shows remarkable accuracy compared to other types of MLIPs.~\cite{Bischoff2024} 
In particular, we trained a MACE MLIP on the Perdew-Burke-Ernzerhof (PBE) functional and used it to study the nature of the LLT as a function of both system size and simulation length. 
This also allows us to directly compare our results with the large body of literature that used the same functional \cite{Bonitz2024,Vorberger2007,Holst2008,Lorenzen2010,Morales2010,Tamblyn2010,Yang2015,Norman2017,Tian2019,Geng2019,Bryk2020,Cheng2020,Karasiev2021,istas2024liquid,Gliaudelis2025,Cheng2021,Bischoff2024}.
Our results reveal a first-order transition between an atomic liquid and a molecular solid at low temperatures ($T < 950$\,K), due to the melting line proximity. At higher temperatures, our simulations indicate a crossover between molecular and atomic liquids in the thermodynamic limit. 

\paragraph*{MACE model from first principles.}
To overcome the time and size limitations of \emph{ab initio} simulations, while also retaining a high accuracy across the LLT, we constructed a MACE MLIP by applying a correction to the loss function usually minimized to find optimal parameters of the model.
During the optimization, the loss and its gradients are evaluated on a subset of training configurations ("batch") of dimension~$N_b$. For the $\mu$-th configuration in the batch, having $N_{\mu}$ atoms, $E_{\mu}^{\textrm{pred}}$, $\mathbf{F}_{\mu, j}^{\textrm{pred}}$, and $\boldsymbol{\sigma}_{\mu}^{\textrm{pred}}$ are the total energy, the force acting on the $j$-th atom and the virial stress tensor, respectively, as predicted by the MLIP while $E_{\mu}^{\textrm{ref}}$, $\mathbf{F}_{\mu, j}^{\textrm{ref}}$, and $\boldsymbol{\sigma}_{\mu}^{\textrm{ref}}$ are the same quantities as calculated with DFT. 
A standard loss function $\mathcal{L}$ is the weighted sum of the mean squared errors (MSEs) of the different observables: 
\begin{align}
    \mathcal{L} & =   w_E \frac{1}{N_{b}} \sum_{\mu = 1}^{N_{b}} \left[ \frac{1}{N_{\mu}}\left( E_{\mu}^{\textrm{pred}}   - E_{\mu}^{\textrm{ref}}\right) \right]^2  \notag\\
    & + w_F \frac{1}{N_{b}} \sum_{\mu = 1}^{N_{b}} \frac{1}{3 N_{\mu}} \sum_{j = 1}^{N_{\mu}} \sum_{\alpha = x ,y ,z }\left[ F^{\textrm{pred}}_{\mu, j, \alpha} - F_{\mu, j , \alpha}^{\textrm{ref}} \right]^2 \notag\\
     & +   w_{\sigma} \frac{1}{N_{b}} \sum_{\mu = 1}^{N_{b}}\frac{1}{9} \sum_{\alpha = x ,y ,z } \sum_{\beta =x , y, z}\left[ \sigma_{\mu, \alpha , \beta}^{\textrm{pred}}  - \sigma_{\mu, \alpha , \beta}^{\textrm{ref}} \right]^2 \label{eq: loss function},
\end{align}
where $w_E$, $w_F$, and $w_{\sigma}$ are tunable weights. 
However, when employing the loss in Eq.~\eqref{eq: loss function} to train the model~\footnote{For building the model, we used $128$ equivariant messages, a correlation order of $3$, and a cutoff radius of $r_c = 3$\,Å}, we noticed the appearance of molecular solid-like structures at high temperatures ($T \sim 1200$\,K) during the dynamics. AIMD simulations at the same conditions and system sizes show how these configurations are not dynamically stable, and, therefore, are an artifact of the model. Even after the inclusion of these structures in the training set, the standard loss function still energetically favored these configurations (see the Supplemental Material (SM) for details~\cite{SM}).
The atomic phase, on the other hand, seemed to be described correctly. 

An analysis of the partial energy error distributions for this model, computed on subsets of the training set describing the 
selected phases, reveals how the loss function $\mathcal{L}$ produces a bias, \emph{i.e.,} a non-zero mean of $E^{\textrm{pred}}_{\mu} -E^{\textrm{ref}}_{\mu}$. Indeed, being a function of the MSEs, $\mathcal{L}$ only guarantees that the total error distribution is centered around zero.

With this in mind, we modified the loss function by including a term that penalizes a global energy bias in the prediction error on the subset of \emph{molecular configurations}, thus improving the description of this phase. Specifically, we used the modified loss $\mathcal{L}' = \mathcal{L} + \Delta \mathcal{L}$ with
\begin{equation}
    \Delta \mathcal{L} = w_E \lambda \frac{1}{N_{\textrm{b}}}\left| \sum_{\mu \in \textrm{mol}} \frac{1}{N_\mu} \left( E^{\textrm{pred}}_{\mu} - E^{\textrm{ref}}_{\mu} \right) \right|, \label{eq: penalty loss function}
\end{equation}
where $\lambda$ is a parameter controlling the relative weight of the penalty with respect to the energy term of the standard loss function, $w_E$ is the same hyperparameter as in Eq.~\eqref{eq: loss function}, and the sum only considers molecular configurations in a given training batch. 
To classify configurations as either atomic or molecular, we used a static criterion that solely depends on atomic positions: A configuration is classified as molecular if the first peak of its radial distribution function $g(r)$ (estimated by fitting the $g(r)$ with a Gaussian function for $r \in [0,1.3]$Å) is larger than a threshold, here set to $1.8$. 

We observed that the models trained using $\mathcal{L}'$ have a much smaller energy error for the molecular solid-like structures, which do not appear anymore during the dynamics at high temperatures. As a consequence of the modified loss, the energy error distribution on the training set is slightly altered, with relative energies between different types of structures in much better agreement with respect to the \emph{ab initio} PBE reference values, as discussed in the SM.

To train the final MLIP, we compiled a dataset of $21\,812$ configurations. 
We extracted $\sim 17\,000$ configurations of $128$ hydrogen atoms each from the MD simulations of Refs.~\cite{Tirelli2022} and \cite{Tenti2024} (see the SM for the density-temperature distribution of this set).
An additional set of $3000$ 128-atoms configurations was selected from a series of AIMD simulations at lower temperatures (800 and 900\,K).
We also added $500$ snapshots with 256 and 512 atoms extracted from MD runs with early iterations of the MACE model. Finally, we included solid structures (see Ref.~\cite{Monacelli2023}) and $\sim 100$ low-temperature $96$-atoms configurations from Ref.~\cite{Niu2023}. 
To ensure consistency across the dataset, we recomputed energies, forces, and pressures at the PBE level, using the \textsc{Quantum Espresso} package~\cite{Giannozzi2009,Giannozzi2017,Giannozzi2020}. A projector augmented-wave (PAW) pseudopotential~\cite{PAWpseudo} together with a $60$~Ry plane-waves cutoff was used. A sufficiently dense $\mathbf{k}$-point grid for each system size was employed to cure finite-size effects. For instance, we used a $4 \times 4 \times 4$ grid for the configurations with $128$ atoms. 

\begin{figure*}[t]
    \centering
    \includegraphics[width=\linewidth]{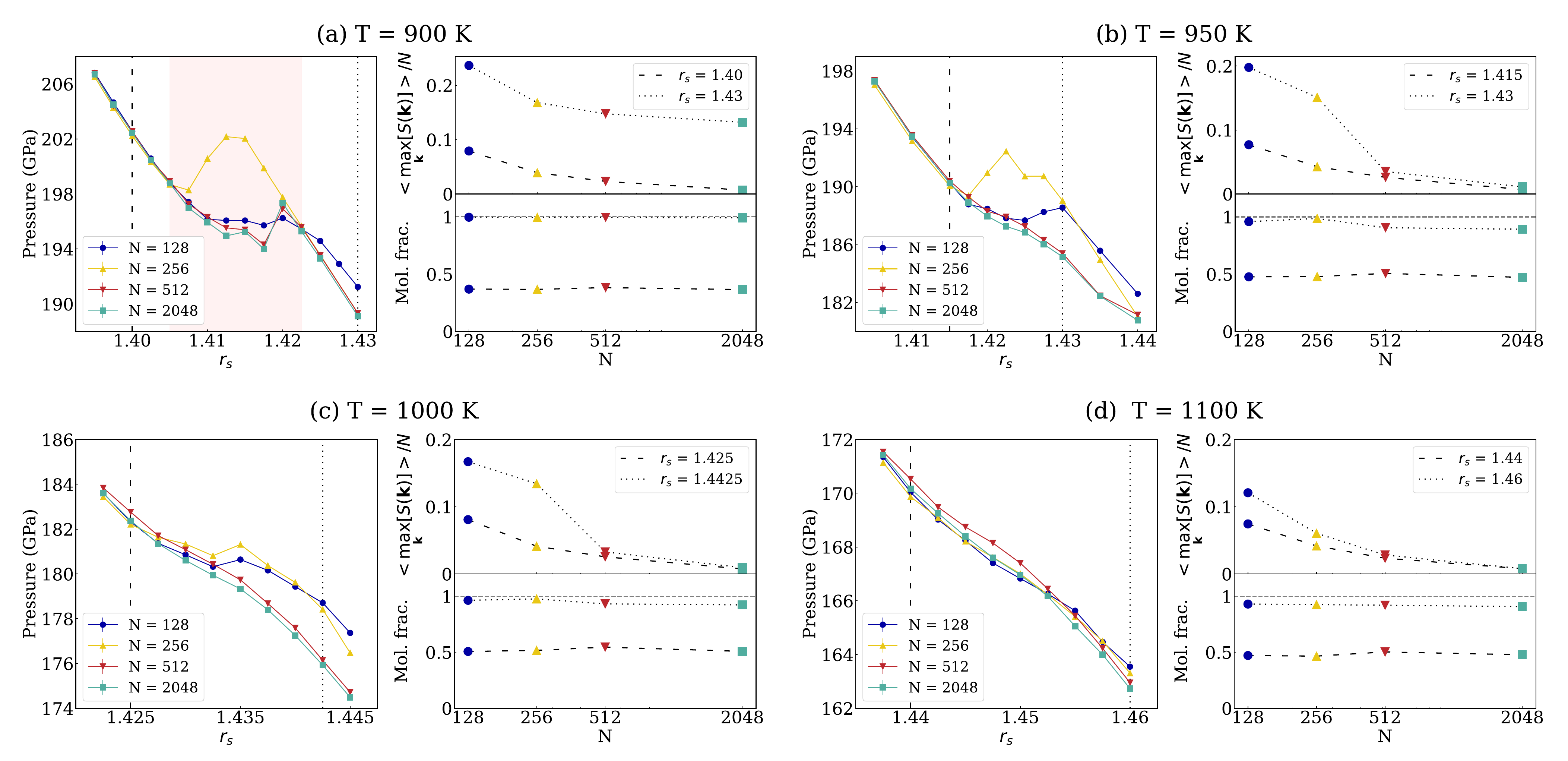}
    \caption{Results for MD simulations obtained with the MACE model for different temperatures: (a) $T=900$~K, (b) $T=950$~K, (c) $T=1000$~K, (d) $T=1100$~K. For each temperature, the left panel reports the EOS close to the LLT for the different system sizes. For $T=900$~K the red shaded area highlights the $r_s$ range for which hysteresis is observed at all values of $N$. The right panels show the average $\max_{\mathbf{k}} S(\mathbf{k}) /N$ along the trajectory (top right) and the average molecular fraction $m$ (bottom right) as a function of the system size $N$ for two different values of the Wigner-Seitz radius $r_s$.}
    \label{fig: dft MACE EOS}
\end{figure*}

From this dataset, we randomly extracted $280$ configurations for testing. From the remaining configurations, $95\%$ were used for training, and $5\%$ for validation. This last group of configurations is used during the training to assess the performance of the model. At the end of the optimization, the best model is selected as the one minimizing the loss on the validation set. 

\begin{table}[h]
    \centering
    \begin{tabular}{l|cc}
    \toprule
        & RMSE$_{E}$ (meV/atom) & RMSE$_F$ (meV/Å)\\
        \midrule
        Training &  $2.2$ & $116$ \\
        Validation & $2.1$ & $118$\\
        Test & $2.0$ & $116$ \\
        \bottomrule
    \end{tabular}
    \caption{Root mean squared errors of the final MACE MLIP for the energy per atom ($\textrm{RMSE}_E$) and the forces ($\textrm{RMSE}_F$) on the training, validation, and test sets.}
    \label{tab: MACE model performances}
\end{table}

Table~\ref{tab: MACE model performances} reports the accuracy of this MACE model, measured by the root mean squared error (RMSE) on the energy per atom and the forces on the training, validation, and test sets. The test-set RMSE for the virial pressure is $\sim1$~GPa. 
Compared to the MLIP proposed in Ref.~\cite{Cheng2020}, our model has an error on energy $6-7$ times smaller and is twice as accurate on forces. The recent NequIP model of Ref.~\cite{istas2024liquid} shows a very similar RMSE on energy, but has a $40\%$ larger RMSE on forces. 

\paragraph*{LLT simulations: finite-size scaling.}\label{sec: LLT simulations: size scaling}
The computational efficiency of the MACE model enabled us to study the behavior of the LLT as a function of both temperature and system size. 
We ran MD simulations in the NVT ensemble using the \textsc{LAMMPS} code~\cite{Thompson2022} interfaced with MACE. 
We considered systems with $N= 128$, 256, 512, and 2048 hydrogen atoms treated as classical nuclei, and cubic supercells. We performed simulations as long as $0.6$\,ns, about two orders of magnitude longer than what is usually achieved with AIMD. For the dynamics, we used a time step of $0.2$~fs and controlled the temperature via the stochastic velocity rescaling scheme~\cite{Bussi2007}, with a characteristic time of $\tau = 0.1$~ps. 

To accurately study the nature of the transition, we calculated the EOS of the system in the vicinity of the LLT, using a dense grid of Wigner-Seitz radii $r_s$, where $\frac{4\pi}{3}\left(r_s a_0\right)^3  = \frac{V}{N}$ with $V$ and $a_0$ being the system volume and the Bohr radius, respectively.
The results are presented in Fig.~\ref{fig: dft MACE EOS} for four different temperatures in the $[900 - 1100]$~K range.
From the EOS, we can identify three distinct regimes for the LLT: 
At the lowest temperature computed here, \ie, $T= 900$~K (Fig.~\ref{fig: dft MACE EOS}a), the model clearly predicts a first-order transition, signaled by the hysteresis present in the EOS for all system sizes.
At $T=950$\,K and $1000$\,K (Figs.~\ref{fig: dft MACE EOS}b and \ref{fig: dft MACE EOS}c, respectively), the EOS has a strong dependence on the system size in the transition region: the smaller systems (\ie, $N = 128$ and 256) suggest again a first-order transition, while for larger $N$ the pressure plateau and hysteresis are missing.
Finally, at $T \geq 1100$~K (Fig.~\ref{fig: dft MACE EOS}d) the results indicate a smooth crossover, with a relatively small size dependence. 

To characterize the LLT, the sole observation of the EOS is not sufficient, since it is known that for these transitions the density is not the primary order parameter~\cite{Tanaka2020}. In particular, we computed the stable molecular fraction~\cite{Tamblyn2010,Geng2019,Bischoff2024} of the system, defined in this case as the average number of hydrogen pairs whose constituent atoms stay within a distance of $1.05$~Å for at least a time $\tau \sim 80$~fs. 

The results corresponding to $r_s$ values slightly before/after the transition are also shown in Fig.~\ref{fig: dft MACE EOS}. In all cases, the molecular fraction rapidly increases from values around $0.5$ to values close to $1$, signaling an atomic-to-molecular transition as $r_s$ increases.

Besides the stable molecular fraction, we also computed the structure factor to detect long-range spatial correlations, \emph{i.e.}, 

\begin{align*}
    S (\mathbf{k}) = \left\langle \frac{1}{N} \sum_{i,j} \exp \bigl( i \mathbf{k}  \cdot \left  ( \mathbf{R}_i - \mathbf{R}_j  \right)\bigr)\right\rangle,
\end{align*}
where $\langle \cdots\rangle $ indicates the average over the trajectory, $\mathbf{R}_1, \dots, \mathbf{R}_N$ are atom positions, and $\mathbf{k} = \frac{2\pi}{L} (n_1 , n_2 , n_3)$, with $L$ being the side of the cubic simulation box and $n_1 , n_2 , n_3$ integer numbers.
In particular, the maximum value of $S (\mathbf{k})$ can be used as a proxy for the formation of crystalline structures, with $\max_{\mathbf{k}} S (\mathbf{k}) \propto N$ in a solid system in the thermodynamic limit.
On the contrary, in a liquid, the rotational invariance implies that the contribution of each $\mathbf{k}$ to the $S(\mathbf{k})$ will stay constant for $N\rightarrow \infty$. 
The value of the average $\max_{\mathbf{k}} S (\mathbf{k}) / N$ as a function of the number of particles is reported in the top right panels of Fig.~\ref{fig: dft MACE EOS}. 
In the atomic phase, our simulations show that the system is a liquid at all temperatures. In the molecular region, the behavior of $\max_{\mathbf{k}} S (\mathbf{k}) / N$ is instead strongly temperature dependent.
Interestingly, the first-order transition observed in the EOS  at $T=900$~K is accompanied by a non-vanishing value of $\max_{\mathbf{k}} S (\mathbf{k}) / N$ in the thermodynamic limit, revealing a long-range spatial order of the molecular phase at this temperature (Fig.~\ref{fig: dft MACE EOS}a). Moreover, at $T= 950$~K and 1000~K, a large value of $\max_{\mathbf{k}} S (\mathbf{k}) / N$ is present for $N=128$ and $256$ on the molecular side, then vanishing at larger $N$ (Figs.~\ref{fig: dft MACE EOS}(b-c)). This suggests that the first-order transition seen in smaller systems is due to finite-size effects and it is between a molecular crystal and an atomic liquid. 
A genuine molecular liquid is instead observed for all system sizes at temperature $T\geq 1100$~K (Fig.~\ref{fig: dft MACE EOS}d).

\paragraph*{LLT simulations: finite-time effects.}
Our finite-size scaling analysis highlights the importance of considering sufficiently large systems to obtain converged results for the LLT. 
\begin{figure}
    \centering  \includegraphics[width=\linewidth]{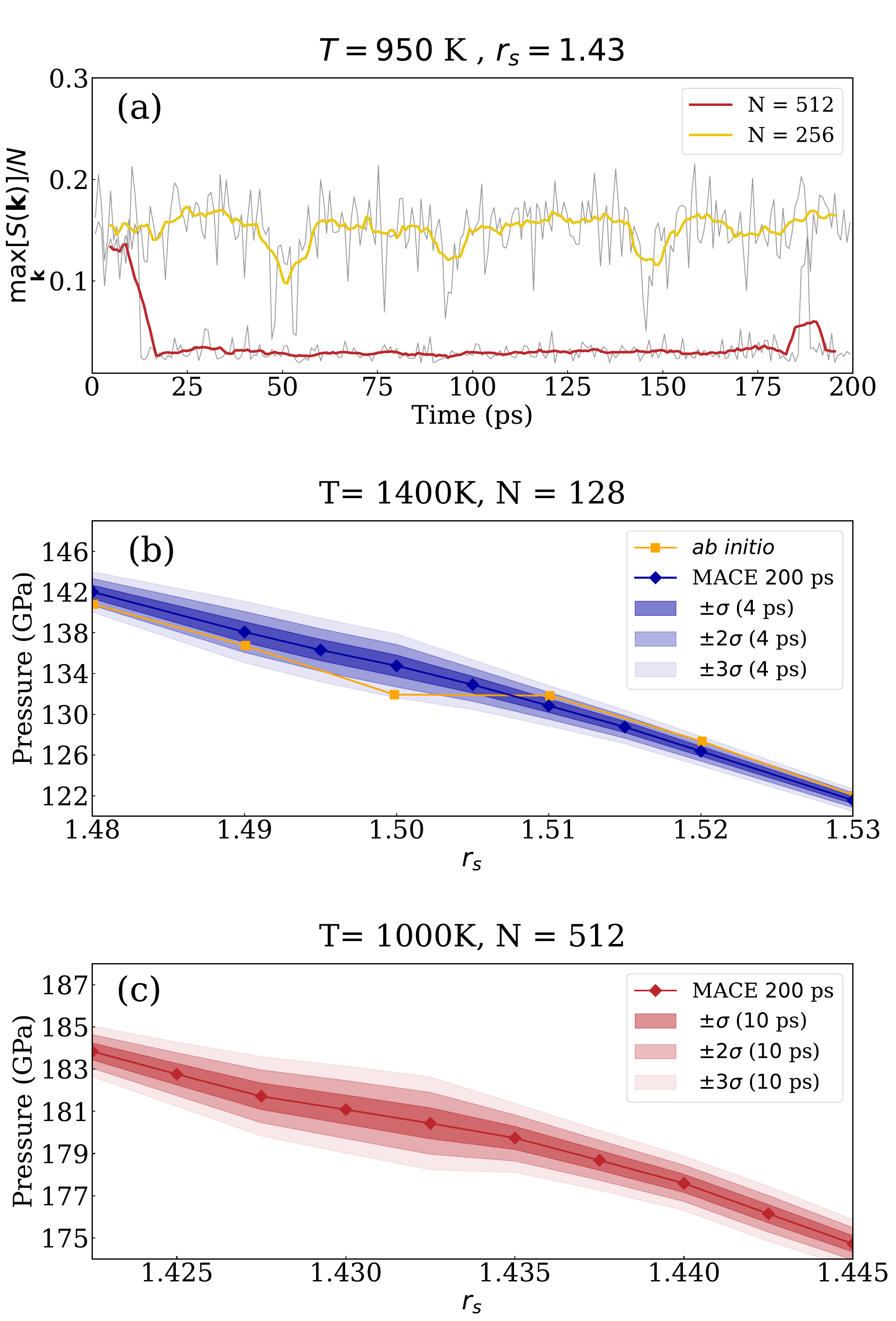}
    \caption{(a) Value of $\max_{\mathbf{k}}S(\mathbf{k}) / N$ as a function of the simulation time for two systems, with $N=256$ and $N=512$ atoms, respectively, at temperature $T= 950$~K and $r_s = 1.43$. Colored lines represent running averages over a time window of $8$~\,ps. (b) Results for $200$~ps long simulations, $N=128$, $T=1400$~K and corresponding confidence intervals, estimated from a running average with time window $\tau_{\textrm{run}} = 4$~ps. The AIMD result reported in Ref.~\cite{Tirelli2022} is also shown. (c) Same as (b) but with $N=512$, $T=1000$~K and $\tau_{\textrm{run}} = 10$~ps.}
    \label{fig: skmaxstory & short length}
\end{figure}
This is further demonstrated in Fig.~\ref{fig: skmaxstory & short length}a, where the behavior of $\max_{\mathbf{k}}S(\mathbf{k}) / N$ is plotted as a function of the simulation time for two sizes, namely $N=256$ and 512, at $T=950$~K and $r_s = 1.43$.
While the smaller system is in a solid-like state for the majority of the time, the system with $N=512$ occasionally crystallizes, as indicated by the two jumps of $\max_{\mathbf{k}}S(\mathbf{k})/N$ at the beginning and the end of the MD run, but mostly remains liquid. 
This behavior not only confirms the size dependence already seen, but also shows that at least $10$ ps are necessary to melt the crystal for $N=512$, suggesting that $\sim 100$ ps are needed to obtain fully converged results near the transition.

\begin{figure}[t]
    \centering
    \includegraphics[width=\linewidth]{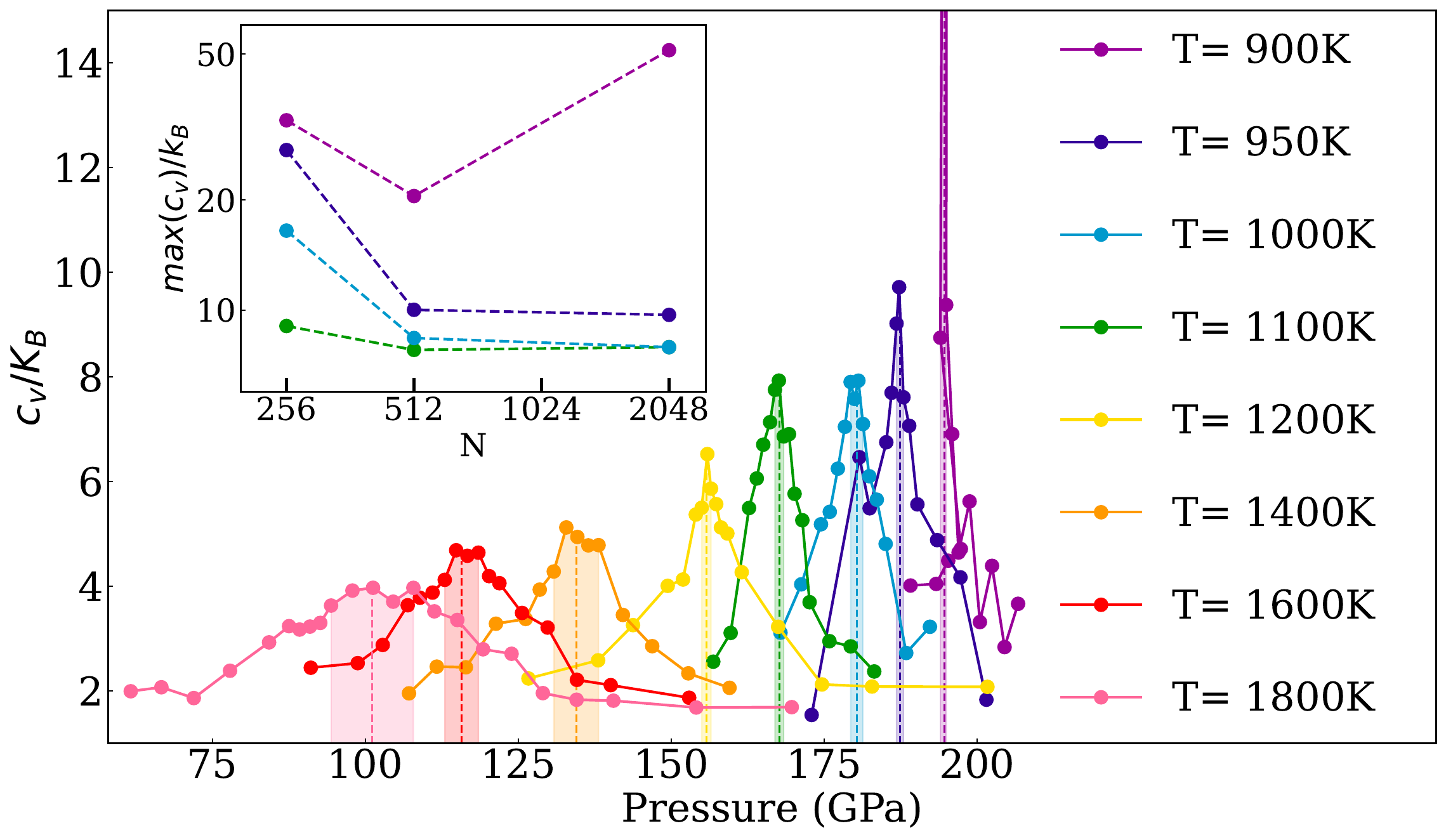}
    \caption{Specific heat per particle vs pressure along the isotherms, for
    $N = 2048$ 
    obtained with the MACE model. The shaded areas indicate the uncertainty in the peak position. In the inset, we report the size scaling of the maximum of $c_v/k_B$ for four different temperatures (in log-log scale). The results confirm a first-order transition for $T=900$~K, where $c_v$ scales with a behavior close to linear, and a smooth crossover for higher temperatures in the thermodynamic limit.}
    \label{fig: CV nvt}
\end{figure}

To analyze the possibility that the kinks identified in the EOS of previous AIMD results originate from the short length of these simulations, we analyzed $0.2$~ns long trajectories obtained with the MACE model for values of $r_s$ close to the transition at different temperatures above our $T_c$. The results are shown in Figs.~\ref{fig: skmaxstory & short length}b and \ref{fig: skmaxstory & short length}c for the system of $N=128$ at $T=1400$~K and the one of $N=512$ at $T=1000$~K, respectively. Using long trajectories, we computed the pressure running average using a variable-size time window $\tau_{\textrm{run}}$, corresponding to simulation times achievable by AIMD, \eg, $\tau_{\textrm{run}} \sim 10$~ps. The variance of the running average gives an idea of the variability of the estimated equilibrium pressure when running MD simulations of length equal to $\tau_{\textrm{run}}$. From Figs.~\ref{fig: skmaxstory & short length}b and \ref{fig: skmaxstory & short length}c, the estimated variance grows in the vicinity of the transition, and the presence of a pressure plateau in the EOS can be understood as an artifact due to the lack of phase-space sampling. 

For the smallest system ($N=128$, Fig.~\ref{fig: skmaxstory & short length}b), we also compared our results with the PBE AIMD ones reported in Ref.~\cite{Tirelli2022} at $T=1400$~K for the same size. The pressure plateau here is within the $3 \sigma$ uncertainty region estimated from a running average of $4$~ps, slightly longer than the average length of the AIMD simulations. 
The results at lower temperature $T=1000$~K for a larger system of $512$ atoms (Fig.~\ref{fig: skmaxstory & short length}c) show that even a $10$~ps long dynamics can, in principle, produce artificial kinks of the size reported in the literature~\cite{Lorenzen2010,Geng2019}. Even though, in principle, this estimation depends on the algorithmic details, such as the choice of the thermostat, we do not expect this to change our conclusions.

\begin{figure}[h]
    \centering
    \includegraphics[width=\linewidth]{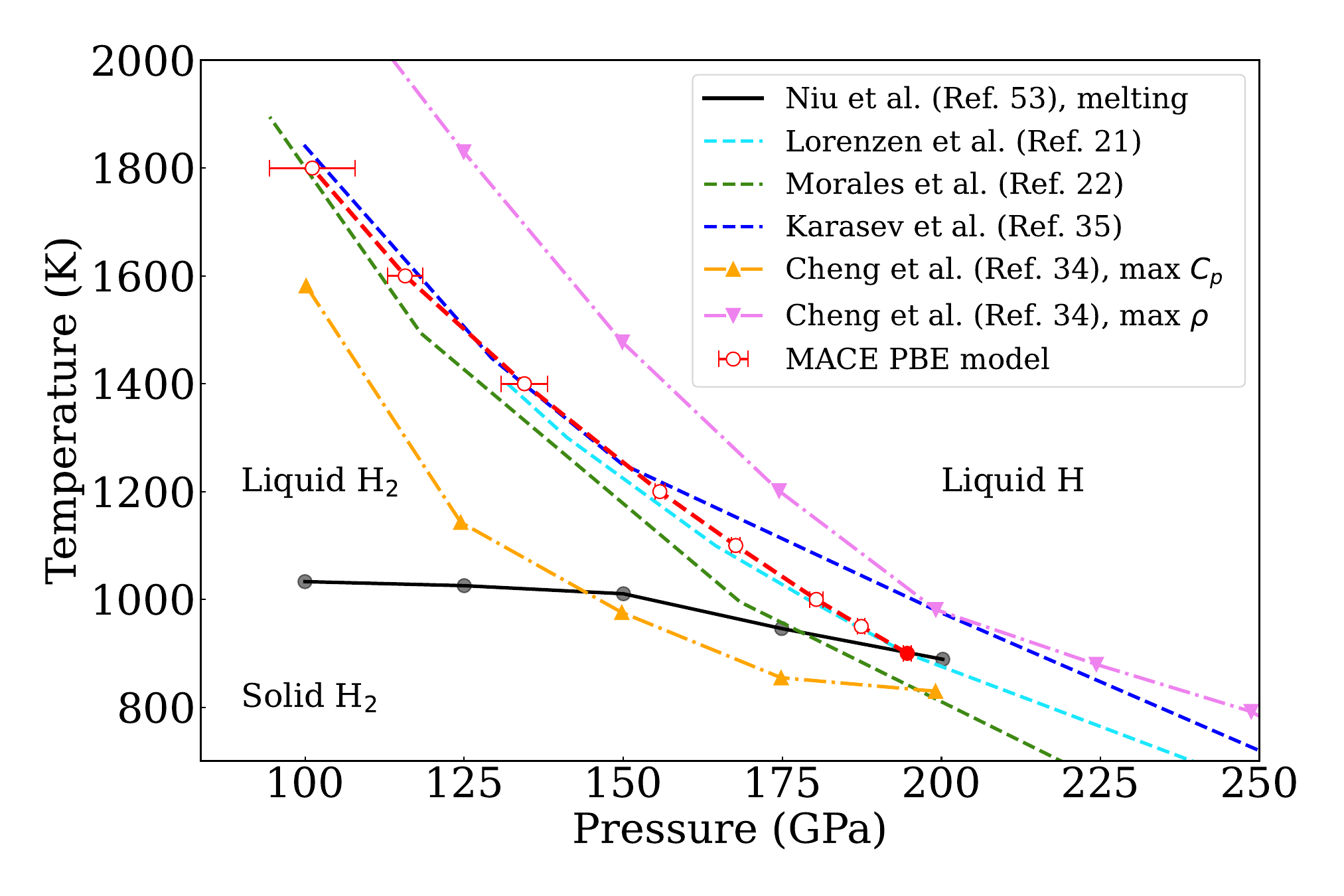}
    \caption{Classical PBE-LLT location as computed with different methods. AIMD results by Ref.~\cite{Lorenzen2010} (light blue dashed line), Ref.~\cite{Morales2010} (green dashed line), Ref.~\cite{Karasiev2021} (blue dashed line). The results for the molecular-to-atomic crossover obtained with an NN MLIP by Ref.~\cite{Cheng2020} are reported with orange and violet markers, corresponding to the maximum of the isobaric specific heat and density, respectively. The black markers and line indicate the recently proposed PBE melting line by Ref.~\cite{Niu2023}, obtained with an NN MLIP and the two-phase method. The MACE model results are indicated with red markers. The filled point at $T= 900$~K indicates the first-order character of the transition, while the empty points correspond to the location of the Widom line given by the $c_v$ maximum (see Fig.~\ref{fig: CV nvt}). }
    \label{fig: LLT PBE}
\end{figure}

\paragraph*{Widom line.}\label{sec: Results for the Widom line}
Both the first-order transition and the crossover above the critical temperature $T_c$ can be further characterized by studying the behavior of the isochoric specific heat (\ie, the heat capacity per particle), $c_v=\frac{1}{Nk_BT^2} ( \left\langle E^2 \right\rangle - \left\langle E \right\rangle^2)$, as a function of temperature, density, and system size.  
The results for $c_v$ obtained from simulations with $N=2048$ hydrogen atoms at temperatures up to $1800$~K are reported in Fig.~\ref{fig: CV nvt}. 
For each temperature, we estimated the location and value of the maximum. For temperatures between $900$~K and $1100$~K, we also studied the size scaling of the $c_v$ maximum, as shown in the inset of Fig.~\ref{fig: CV nvt}. In a first-order transition, the $c_v$ peak presents a divergence scaling like $\sim N$ in the thermodynamic limit~\cite{Binder1987}. As shown in the figure, linear scaling is observed at $T=900$~K, while $c_v$ tends to a constant for higher temperatures. This is consistent with our EOS results locating the critical point between $T=900$~K and $T=950$~K. 
The position of the $c_v$ maximum at higher temperatures allows one to determine the Widom line in the supercritical region. The results are reported in Fig.~\ref{fig: LLT PBE}. 
Our Widom line shows remarkable agreement with previous simulations of the LLT obtained with the PBE functional. In particular, our results well reproduce the ones reported in Ref.~\cite{Lorenzen2010} for temperatures up to $1400$~K and those of Ref.~\cite{Karasiev2021} for higher temperatures, even though they do not agree on the LLT first-order character for $T\geq 900$~K. 

\paragraph*{Conclusions.}
Enabled by an accurate and efficient MLIP based on the MACE architecture, we show that the LLT between molecular and atomic hydrogen is always supercritical above the melting temperature.
For this, we performed NVT MD simulations in a temperature range between $900$\,K and $1800$\,K of systems with up to $2048$ hydrogen atoms and simulation times of up to $600$\, ps.  
We preserved the \emph{ab initio} PBE quality of the MACE model across the transition by virtue of a modified loss function, yielding consistent energy differences between atomic and molecular configurations.

For temperatures above $900$\,K, the MACE model predicts a proper LLT in the thermodynamic limit ($N\geq 512$), which turns out to be a smooth crossover. 
Up to $T=1000$\,K, defective crystal structures are observed for small systems due to strong finite-size effects, giving rise to a first-order transition that disappears as $N$ increases.

At the lowest temperature ($900$\,K), the model predicts a first-order transition in the thermodynamic limit. The structure factor analysis reveals the presence of long-range spatial order in the molecular phase, showing that this is a transition between a molecular solid-like system, frustrated by the cubic cell, and an atomic liquid. 
This transition point lies exactly on the PBE melting line of Ref.~\cite{Niu2023} (see Fig.~\ref{fig: LLT PBE}), calculated with an NN MLIP obtained using the DeePMD package~\cite{Zeng2023}, suggesting that the LLT might become first-order inside the crystal region for $T\leq 900$~K.

The physical picture provided by the MACE model qualitatively agrees with Ref.~\cite{Cheng2020} on the supercritical nature of the LLT, even though their model could not accurately reproduce the crystallization regime, and predicted a Widom line (estimated from the maximum of the isobaric specific heat~$c_p$) far from the AIMD results (see Fig.~\ref{fig: LLT PBE}).
The first-order character of the transition observed in other studies may be explained by the large fluctuations expected close to the Widom line, which could lead to the incorrect identification of density jumps and pressure plateaux in the EOS for too short simulation times. 
Additional path integral MD simulations carried out with the MACE model indicate that the conclusions about the crossover nature of the LLT are not changed by the inclusion of nuclear quantum effects, as reported in the SM~\cite{SM}.

These results urgently call for a reinvestigation of the LLT using \emph{ab initio} methods beyond DFT, such as QMC. The $\Delta$-learning scheme~{\cite{Tirelli2022, Tenti2024}} combined with the present MACE model taken as baseline could allow one to access unprecedented system sizes and simulation lengths, by extending this work to QMC-based MLIP applications.

\paragraph*{Data availability.}
A version of the MACE code implementing the modified loss function employed in this work can be found at Ref.~\cite{github}. The model, training set, and simulation results are available at Ref.~\cite{zenodo}. 

\paragraph*{Acknowledgments.}
K.N. acknowledges financial support from MEXT Leading Initiative for Excellent Young Researchers (Grant No.~JPMXS0320220025) and from by JST BOOST (Grant No.~JPMJBY24F3). B.J. acknowledges support by the state of Baden-Württemberg through bwHPC. M. C. thanks the European High Performance Computing Joint Undertaking (JU) for the partial support through the "EU-Japan Alliance in HPC" HANAMI project 
(Hpc AlliaNce for Applications and supercoMputing Innovation: the Europe - Japan collaboration). 
This work has received funding from the European Center of Excellence in Exascale Computing TREX (Targeting Real chemical accuracy at the Exascale, grant 952165).
The authors acknowledge discussions with Prof.~David Ceperley, Prof.~Michele Ceriotti, Prof.~Gábor Csányi, Prof.~Carlo Pierleoni, Dr.~Thomas Bischoff, Dr.~Marco Cherubini, and Dr.~Abhishek Raghav.
M.C. acknowledges GENCI for providing computational resources on the CEA-TGCC Irene and IDRIS Jean-Zay supercomputer clusters under project number A0170906493 and the TGCC special session.

\bibliography{Bibliography}
\end{document}


\title{Supplemental Material for ``Hydrogen liquid-liquid transition from first principles and machine learning"}

\author{Giacomo Tenti} 
\email{gtenti@sissa.it}
\affiliation{International School for Advanced Studies (SISSA),
Via Bonomea 265, 34136 Trieste, Italy}

\author{Bastian Jäckl} 
\affiliation{Department of Computer and Information Science, University of Konstanz, Box 188, 78457 Konstanz, Germany}

\author{Kousuke Nakano} 
\affiliation{Center for Basic Research on Materials, National Institute for Materials Science (NIMS), 1-2-1 Sengen, Tsukuba, Ibaraki 305-0047, Japan}
\affiliation{Center for Emergent Matter Science (CEMS), RIKEN, 2-1 Hirosawa, Wako, Saitama, 351-0198, Japan.}

\author{Matthias Rupp} 
\affiliation{Department of Computer and Information Science, University of Konstanz, Box 188, 78457 Konstanz, Germany}
\affiliation{Scientific Instrumentation and Process Technology unit, Luxembourg Institute of Science and Technology (LIST), Maison de l'Innovation, 5 avenue des Hauts-Fourneaux, L-4362 Esch-sur-Alzette, Luxembourg}

\author{Michele Casula}
\email{michele.casula@impmc.upmc.fr}
\affiliation{Institut de Minéralogie, de Physique des Matériaux et de Cosmochimie (IMPMC), Sorbonne Université, CNRS UMR 7590, MNHN, 4 Place Jussieu, 75252 Paris, France}

\date{\today}
\maketitle

\section{MACE model training details}

\subsection{PBE dataset density-temperature distribution}
\begin{figure}[h]
    \centering
    \includegraphics[trim = {6cm 0 8cm 0}, width=0.67\linewidth]{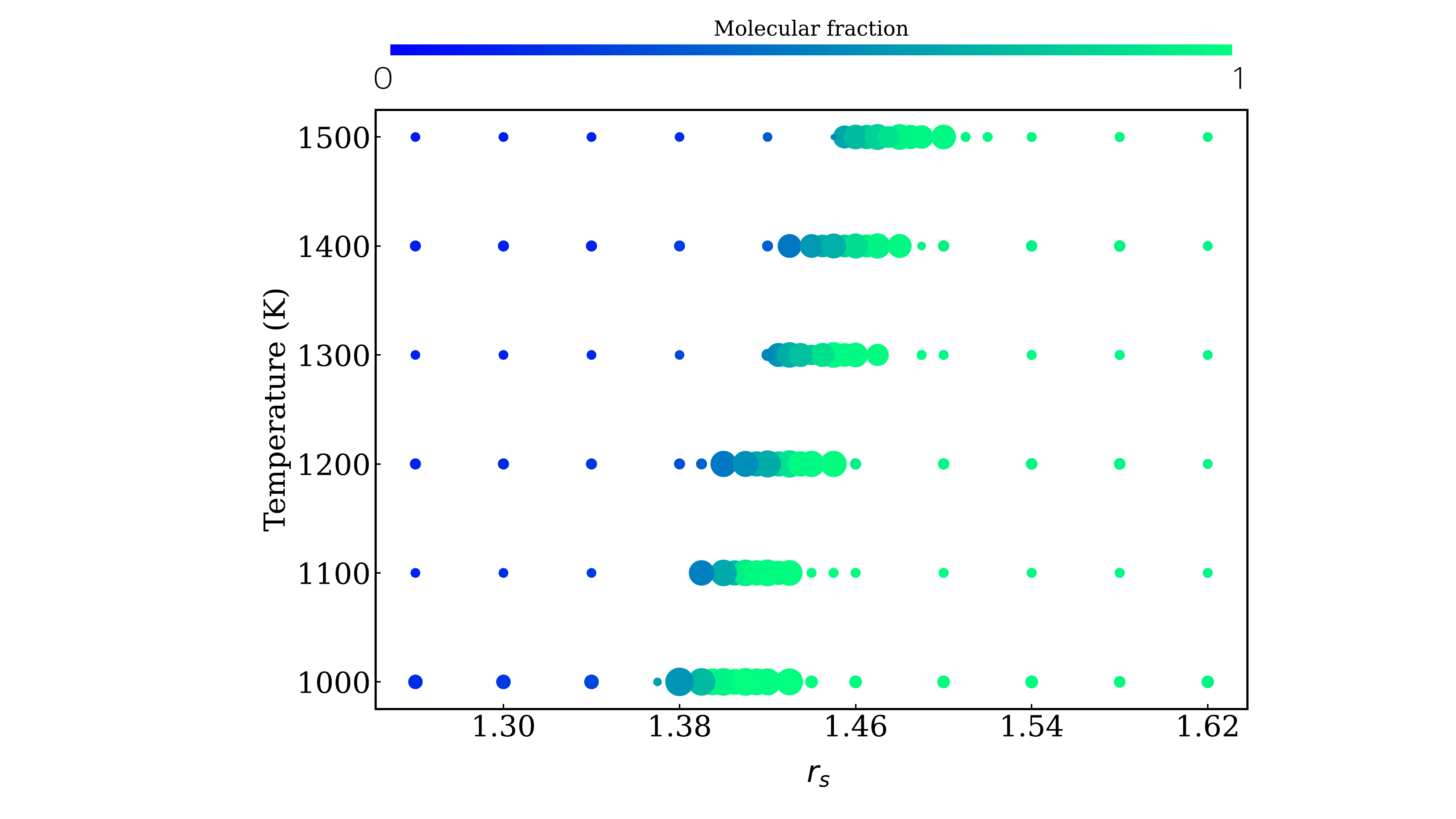}
    \caption{Distribution of configurations sampled from MD runs of Ref.~\cite{Tirelli2022} to generate part of the dataset of the MACE model. The dimension of the dots is proportional to the number of configurations extracted from each MD simulation, while the color indicates the stable molecular fraction. }
    \label{fig: dataset MACE from Tirelli2022}
\end{figure}

As mentioned in the main text, the dataset used for both the model training and validation is a combination of configurations extracted from multiple sources~\cite{Tirelli2022,Tenti2024,Niu2023} and from other trajectories obtained with both {\it{ab initio}} molecular dynamics (AIMD) and dynamics using early iterations of the model. A majority of configurations ($\sim 17000$) were sampled from variation Monte Carlo-based trajectories of Ref.~\cite{Tirelli2022}. This set spans an $r_s$ range from $1.26$ to $1.62$, and temperatures from $T=1000$~K to $T=1500$~K. The dataset comprises both molecular and atomic configurations, as we can observe by computing the stable molecular fraction (as defined in the main text), with an H$_2$ lifetime threshold of $20$~ps. The distribution of the set in the $r_s-T$ space is reported in Fig.~\ref{fig: dataset MACE from Tirelli2022}.

\subsection{Traning parameters}
For the final MACE model, we used $128$ equivariant messages, a correlation order of $3$, two message-passing layers, and a cutoff radius of $r_c = 3$\,Å.
The training was performed using the Adam optimizer~\cite{Kingma2014} with a $N_b = 16$ batch size and initial learning rate of $0.01$, and using the modified loss function $\mathcal{L}'$ defined in the main text.
The loss weights were taken equal to $w_E = 1$, $w_F = 100$, $w_{\sigma} = 100$ for the first $320$ epochs (defined as $N_t / N_b$ optimization steps, with $N_t$ being the total training set dimension) and $w_E = 200$, $w_F = 10$, $w_{\sigma} = 10$ for the remaining $130$ epochs. This is done to reduce the error on the energy after the forces are converged. The relative weight of the energy penalty $\Delta \mathcal{L}$ (see Eq.~2 of the main text) was set to $\lambda = 50$.

\subsection{Modified loss effect on the error distribution}\label{subsec: Modified loss effect on the error distribution}

The effect of using the modified loss function was studied by comparing the energy error distribution $\frac{1}{N_{\mu}}(E^{\textrm{pred}}_{\mu} - E^{\textrm{ref}}_{\mu})$ across the training set for two different models. In particular, we considered the final MACE model employed for the simulations (using the modified loss), and a model trained with a standard loss function (Eq.~1 of the main text) and the same training parameters. 
Moreover, we split the training set into three partitions:
\begin{enumerate}
    \item Solid-like configurations extracted from molecular dynamics (MD) simulations with previous iterations of the model (trained using a standard loss); 
    \item molecular configurations not belonging to the first 
    partition
    (both solid and liquid); 
    \item liquid atomic configurations. 
\end{enumerate}

\begin{figure}[h!]
    \centering
    \includegraphics[width=0.9\linewidth]{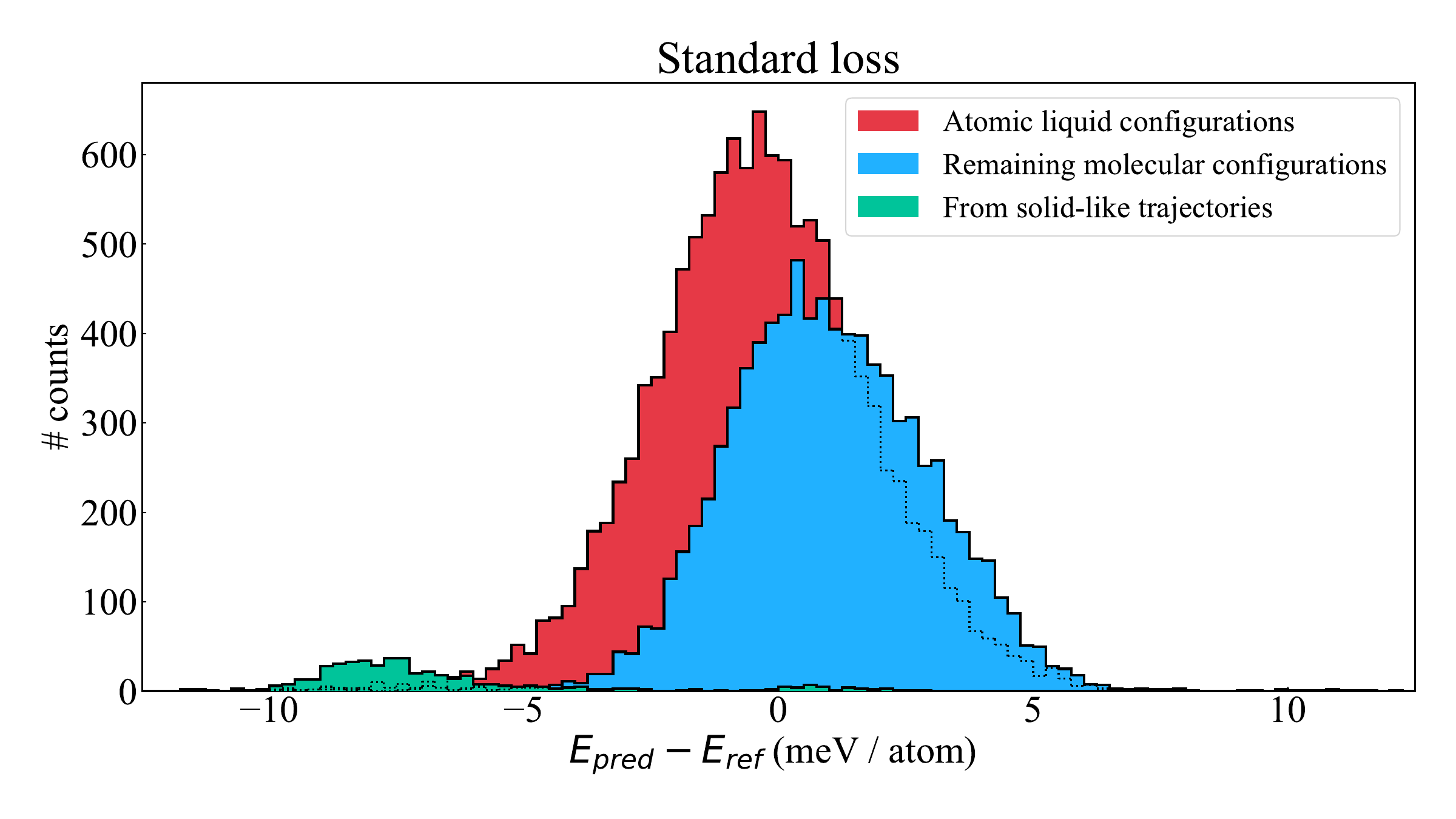}
    \caption{Histogram of the difference between the energy per atom predicted by a model trained using a "standard" loss and the reference PBE value for different partitions of the training set.}
    \label{fig: std loss error distribution}
\end{figure}

\begin{figure}[h!]
    \centering
    \includegraphics[width=0.9\linewidth]{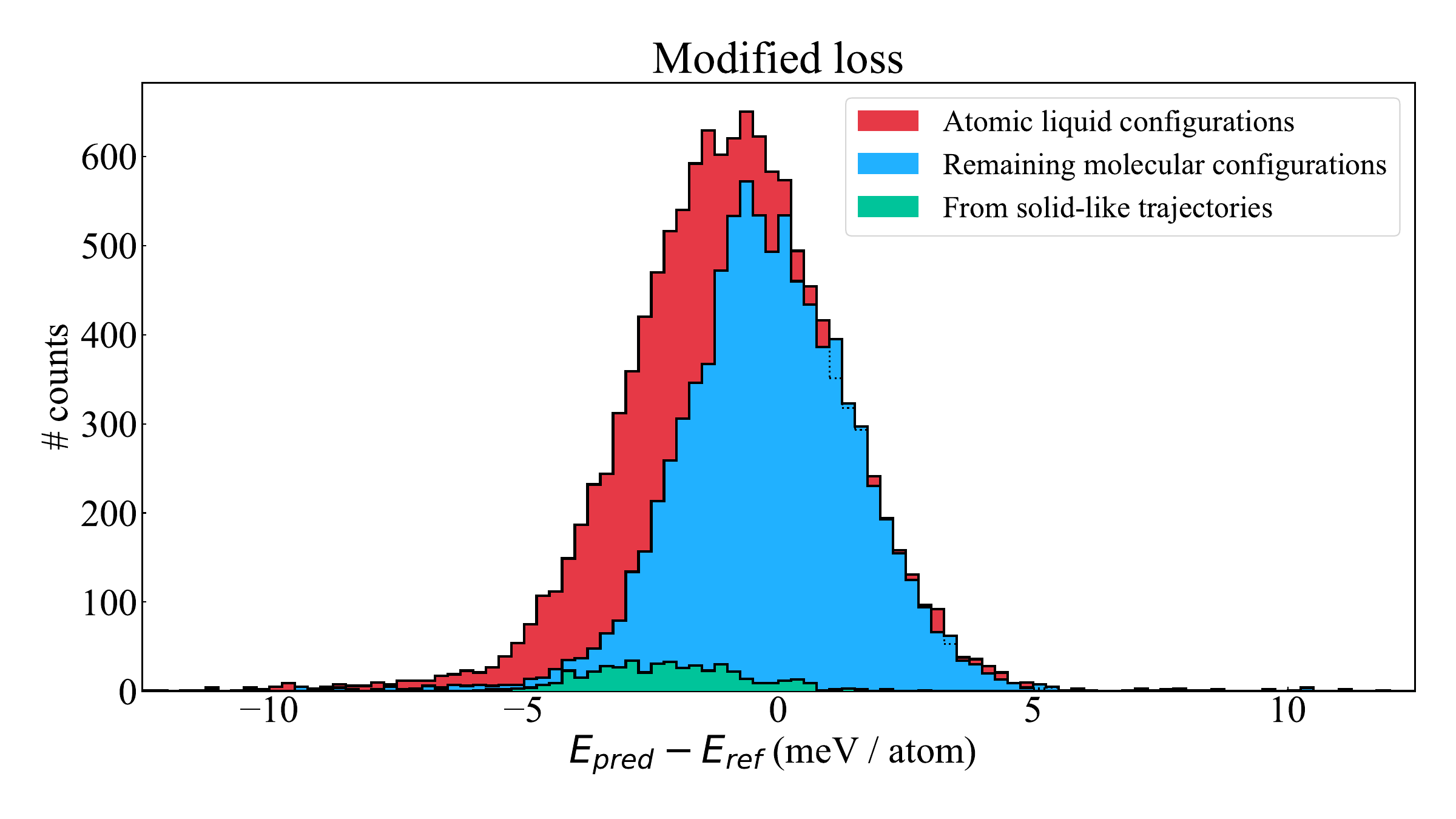}
    \caption{Histogram of the difference between the energy per atom predicted by a model trained using a "modified" loss (introduced in the main text) and the reference PBE value for different partitions of the training set.}
    \label{fig: mod loss error distribution}
\end{figure}

The classification between molecular and atomic configurations followed the static criterion used in the definition of $\Delta \mathcal{L}$, based on the analysis of the radial distribution function (RDF) first peak. 
The results for the error distributions are shown in Figs.~\ref{fig: std loss error distribution} and \ref{fig: mod loss error distribution} for the standard and modified loss, respectively. In Tab.~\ref{tab: rmse-mae-delta std vs mod loss} we also report, for each model and training set partition, the values for the root mean squared error (RMSE), the mean absolute error (MAE) and the average energy shift: 
\begin{align*}
    \Delta_E = \frac{1}{|S|}\sum_{\mu \in S} \frac{1}{N_{\mu}} \left( E^{\textrm{pred}}_{\mu} - E^{\textrm{ref}}_{\mu}\right), 
\end{align*}
where $S$ is one of the training set partitions and $|S|$ is the number of configurations belonging to it. 

\begin{table}[h]
    \centering
    \begin{tabular}{l| ccc | ccc |ccc}
    \toprule
    & \multicolumn{3}{c|}{Atomic liquid} & \multicolumn{3}{c|}{Solid-like} & \multicolumn{3}{c}{Remaining molecular}  \\
    \cline{2-10}
    & RMSE$_E$ & MAE$_E$ & $\Delta_E$ & RMSE$_E$~ & MAE$_E$ & $\Delta_E$ & RMSE$_E$ & MAE$_E$ & $\Delta_E$\\
    \midrule
    Standard & $2.2$ & $1.7$ & $-0.3$ & $7.2$ & $6.8$ & $-6.6$ & $2.6$ & $1.8$ & $0.9$\\
    Modified & $2.3$ & $1.8$ & $-0.9$ & $2.6$ & $2.3$ & $-2.2$ & $1.9$ & $1.4$ & $-0.2$ \\
    \bottomrule
    \end{tabular}
    \caption{Values of the energy RMSE, MAE and shift $\Delta_E$ for the different training set partitions as obtained with two models trained with the standard and modified loss, respectively. All values are reported in meV/atom.}
    \label{tab: rmse-mae-delta std vs mod loss}
\end{table}
We can immediately notice that the standard model dramatically underestimates the energy of the first partition, thus explaining the appearance of these configurations during the dynamics even at relatively high temperatures ($T \sim 1200$~K). We recognized these configurations as unphysical since they do not appear in AIMD, contrary to the solid-like configurations described in the main text, which are at the origin of the first-order transition at $T=900$~K. The use of the modified loss strongly improves the energy of this partition, decreasing the bias towards negative energies. 
The analysis of the error distribution of the other two partitions is also interesting. The RMSE, MAE and absolute value of the shift $\Delta_E$ relative to the "remaining molecular'' partition are all reduced in the modified model with respect to the standard one. This is not surprising, since the extra term in the loss function $\Delta \mathcal{L}$ is directly proportional to $\Delta_E$ evaluated on the molecular configurations. On the contrary, a larger absolute value of the bias is observed in the modified model for the "atomic liquid" partition, even though the two models have almost the same RMSE and MAE. Moreover, notice how the relative difference between the energy shift on the molecular and atomic partitions is overall smaller in the model using the modified loss. We can expect this quantity to be crucial to accurately describe the PBE liquid-liquid phase transition (LLT), so that the modified model should necessarily be better for this task. 


Finally, we mention that we tested different forms for the energy penalty $\Delta \mathcal{L}$, also including an atomic term, e.g., 
\begin{align*}
    \Delta \mathcal{L} = w_E \lambda \frac{1}{N_{\textrm{b}}} \left\{ \left| \sum_{\mu \in \textrm{mol}} \frac{1}{N_\mu} \left( E^{\textrm{pred}}_{\mu} - E^{\textrm{ref}}_{\mu} \right) \right| + \left| \sum_{\mu \in \textrm{atom}} \frac{1}{N_\mu} \left( E^{\textrm{pred}}_{\mu} - E^{\textrm{ref}}_{\mu} \right) \right|\right\} \label{eq: penalty loss function}. 
\end{align*}

The resulting model showed a similar accuracy, but was overall harder to optimize. In the end, we chose to use a correction depending only on the molecular energy shift.

\section{Solid-like structures analysis}

\begin{figure}[h]
    \centering
    \includegraphics[width=1.0\linewidth]{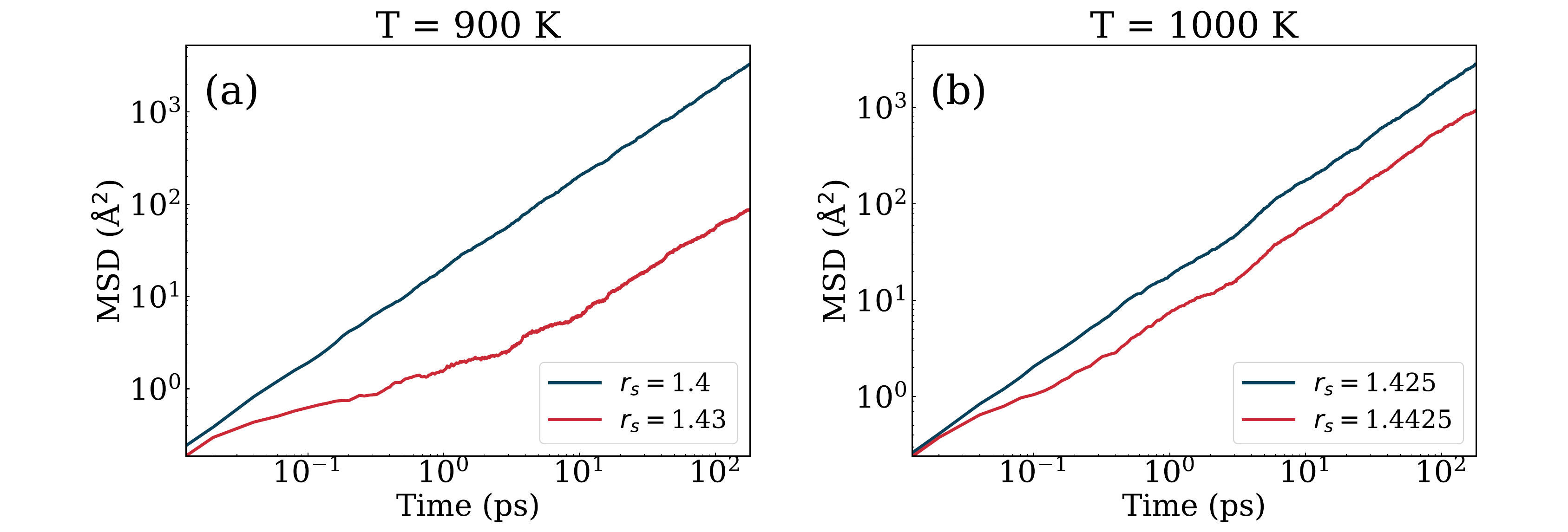}
    \caption{MSD as a function of time for (a) $T=900$~K and $r_s = 1.40$, $r_s = 1.43$ (b) $T=1000$~K and $r_s = 1.425$, $r_s = 1.4425$.}
    \label{fig: MSD model}
\end{figure}

To further characterize the solid-like configurations at the lowest temperature $T=900$~K, we computed the mean squared displacement (MSD) along the trajectory:
\begin{equation}
    \textrm{MSD}(t) = \frac{1}{N} \sum_{i =1}^N \left|\mathbf{R}_i (t) - \mathbf{R}_i (0)\right|^2 . 
\end{equation}
The MSD computed for $T=900$~K and $T=1000$~K for different values of $r_s$ is shown in Fig.~\ref{fig: MSD model}. Notice how the "solid-like" system (corresponding to $T=900$ and $r_s = 1.43$) shows a non-zero diffusivity. In Ref.~\cite{Karasiev2021}, this observation was used as an indication of the liquid phase. Here we obtained very similar values of the MSD (see the supplementary material of Ref.~\cite{Karasiev2021}), although our measure of the $\max_{\mathbf{k}}S(\mathbf{k}) / N$ clearly indicates the presence of long-range spatial correlations.

\begin{figure}[h!]
    \centering  \includegraphics[width = 0.75\textwidth, trim={0cm 0cm 0cm 0cm},clip ]{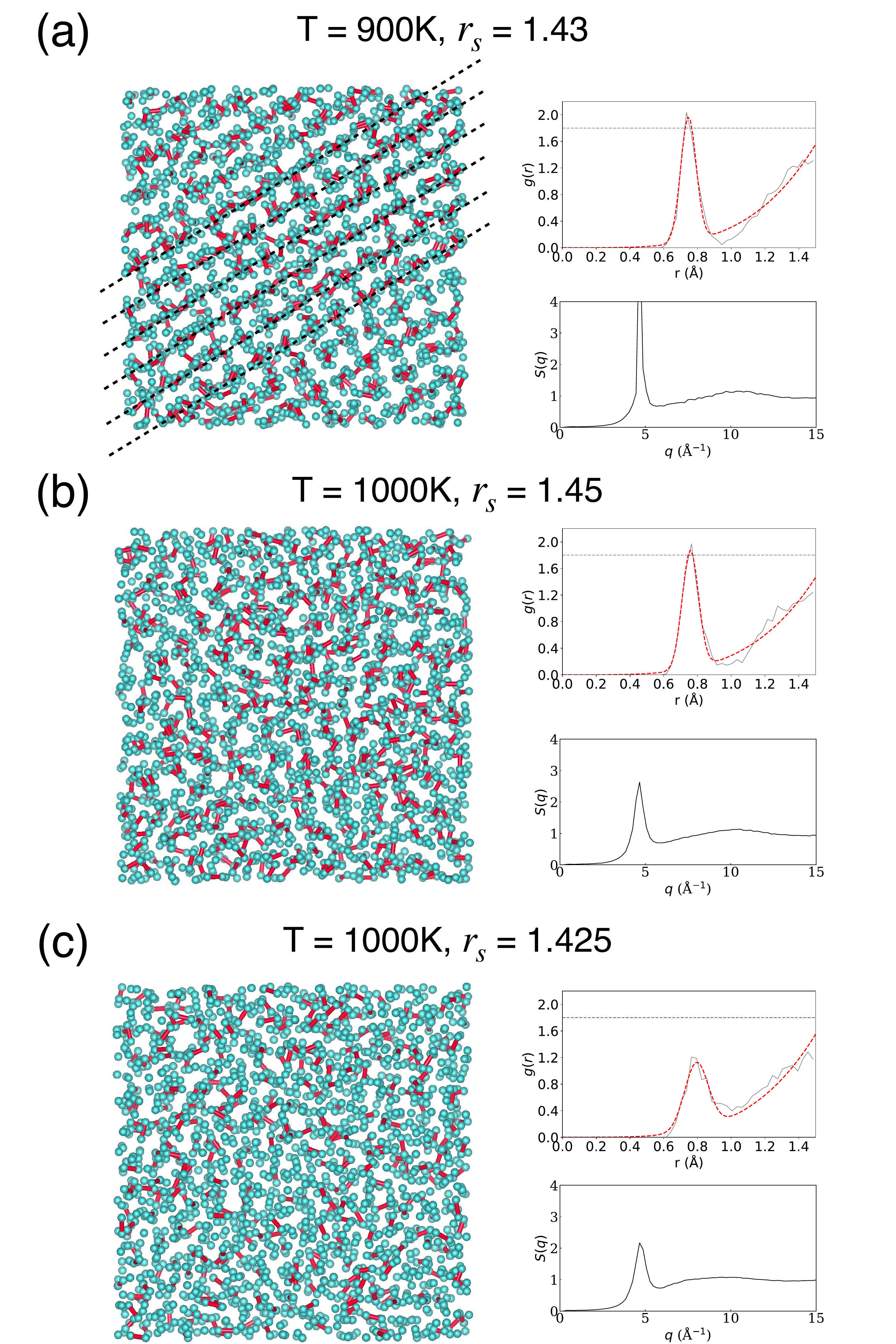}
    \caption{$2048$ atoms structures extracted from classical MD simulations at different temperatures and $r_s$ values: (a) $T=900$~K and $r_s = 1.43$ (molecular crystal); (b) $T=1000$~K and $r_s = 1.45$ (molecular liquid); (c) $T=1000$~K and $r_s = 1.425$ (atomic liquid). The dashed lines in panel (a) are a guide for the eye to highlight the intermolecular planes. For each structure, we report the corresponding radial distribution function (RDF) (gray line), together with the approximant (red line) employed for the classification of the molecular environments in the modified loss function (the threshold of $1.8$ on the RDF peak is also shown). The spherically averaged structure factor $S(q)$ is also depicted. The left images were obtained using the \textsc{VESTA} visualization program~\cite{Momma2008}. Bonds are shown between atoms at a distance $< 0.8$~Å. }
    \label{fig: solid structure}
\end{figure}

The structure of these configurations can be appreciated in Fig.~\ref{fig: solid structure}(a),  where we show a snapshot taken from an MD simulation at $T=900$~K and $r_s = 1.43$ for a $2048$ atoms system. The presence of "planes" of hydrogen molecules in the system can be noticed. The analysis of the $S(\mathbf{k})$ reveals multiple groups of non-collinear reciprocal vectors $\pm \mathbf{k}$ with similar contributions (and scaling with $N$), confirming the solid nature of this phase. 
In Fig.~\ref{fig: solid structure}(b) and \ref{fig: solid structure}(c), we also report structures at higher temperature $T = 1000$\,K in both the molecular and atomic phase. The radial distribution function and spherically averaged structure factor $S(q)$ are also shown to highlight their difference. The atomic snapshots images are obtained using VESTA~\cite{Momma2008}.

To investigate the underlying crystal symmetry, we performed DFT structural optimization of the system using the \textsc{Quantum Espresso} package. In particular, we relaxed the atomic coordinates and cell under a constant pressure of $p = 200$~GPa. Analysis of the final structure reveals a Pbca-16 symmetry. We compared the final ground state DFT-PBE energy per atom of this system with the ones obtained considering a Fmmm-4 structure (proposed in Ref.~\cite{Niu2023} as the correct solid symmetry near the melting line) and a C2c-24 structure, which should well represent the correct crystal at $T=0$\,K for this pressure. Also for these other symmetries the structure was relaxed at $p = 200$~GPa. We computed the energies for each structure with both the modified MACE model and a model trained on a standard loss function (see Sec.~\ref{subsec: Modified loss effect on the error distribution}). The modified model predicts a higher energy for both the Fmmm-4 and Pbca-16 structures, while correctly reproducing their near degeneracy. In principle, this can lead to melting via structural frustration. On the contrary, the model trained with the standard loss function has a smaller error for the Fmmm-4 symmetry, but does not capture the energy degeneracy of the two structures. Finally, we note a higher accuracy of the modified model for the C2c-24 crystal.

\begin{table}[h]
    \centering
    \begin{tabular}{l|ccc}
    \toprule
    & \multicolumn{3}{c}{Energy per atom (eV/atom)}\\
     \cline{2-4} 
    & PBE & Standard Loss & Modified Loss\\
    \midrule
    C2c-24     &  $-14.625$ & $-14.611$  & $-14.624$\\
    Fmmm-4     &  $-14.584$ & $-14.582$  & $-14.568$\\
    Pbca-16    &  $-14.585$ & $-14.568$  &  $-14.569$\\
    \bottomrule
    \end{tabular}
    \caption{Energy per atom for different solid phase symmetries as computed with {\it{ab initio}} PBE, and the models trained with a standard loss function and a modified one, respectively. }
    \label{tab: solid phases GS energy}
\end{table}

\clearpage
\section{PIMD results}
\begin{figure}
    \centering
    \includegraphics[width=0.6\linewidth]{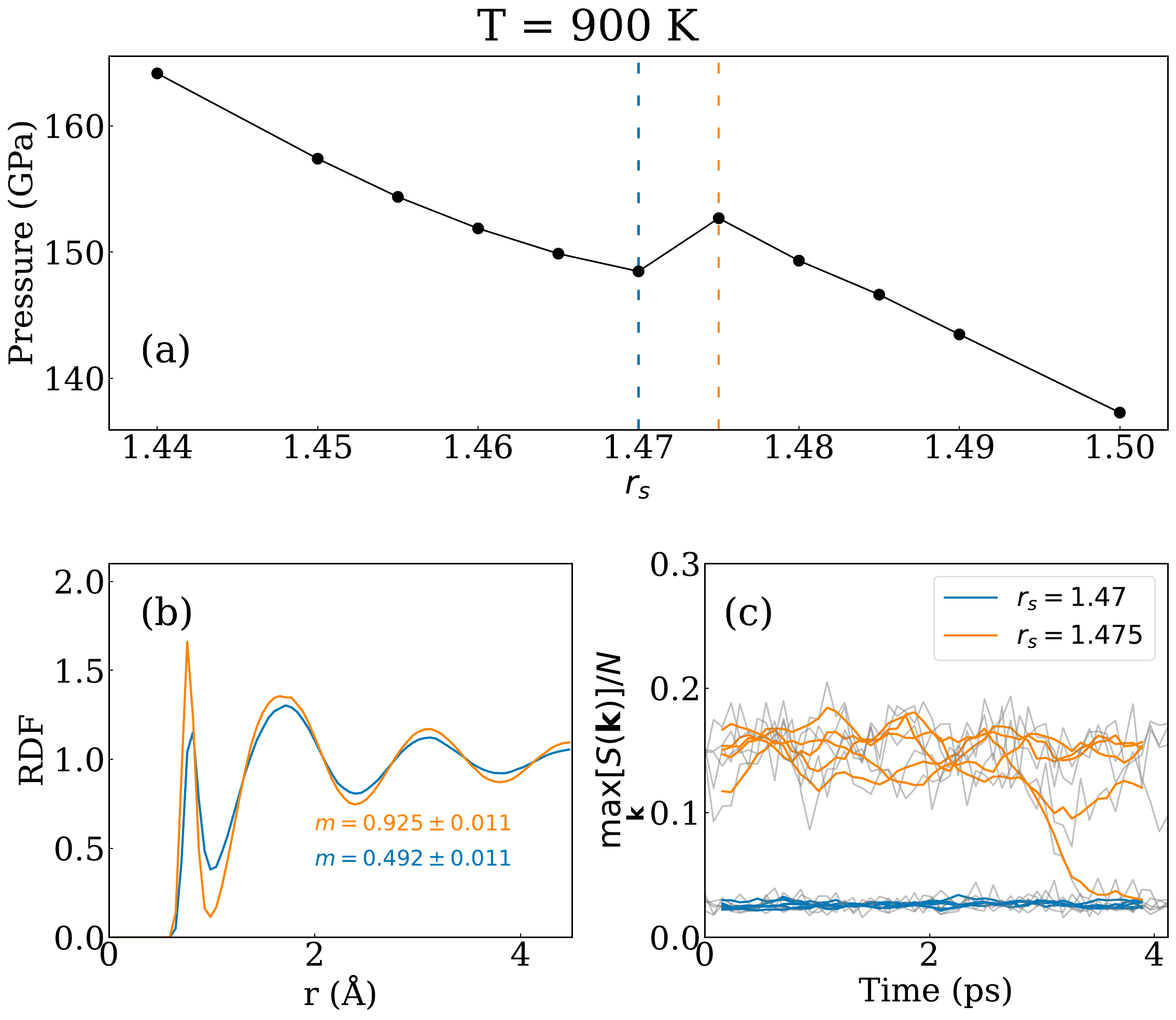}
    \caption{Results of PIOUD simulations on a $512$ hydrogen atoms system at $T = 900$\,K. (a) Equation of state close to the LLT. The vertical dashed lines highlight the $r_s$ values for which we show additional analysis. (b) Radial distribution function for the two $r_s$ values, i.e.,  $r_s = 1.47$ (blue) and $r_s = 1.475$ (orange). The value of the molecular fraction $m$ (as defined in the main text) is also shown. (c) Value of $\max_{\mathbf{k}} S(\mathbf{k}) /N $ along the trajectories for the two $r_s$ values and its running average on a time window of $375$\,fs. The different lines with the same color correspond to independent runs starting with different initial conditions. }
    \label{fig: PIMD T = 900K}
\end{figure}

\begin{figure}
    \centering
    \includegraphics[width=0.6\linewidth]{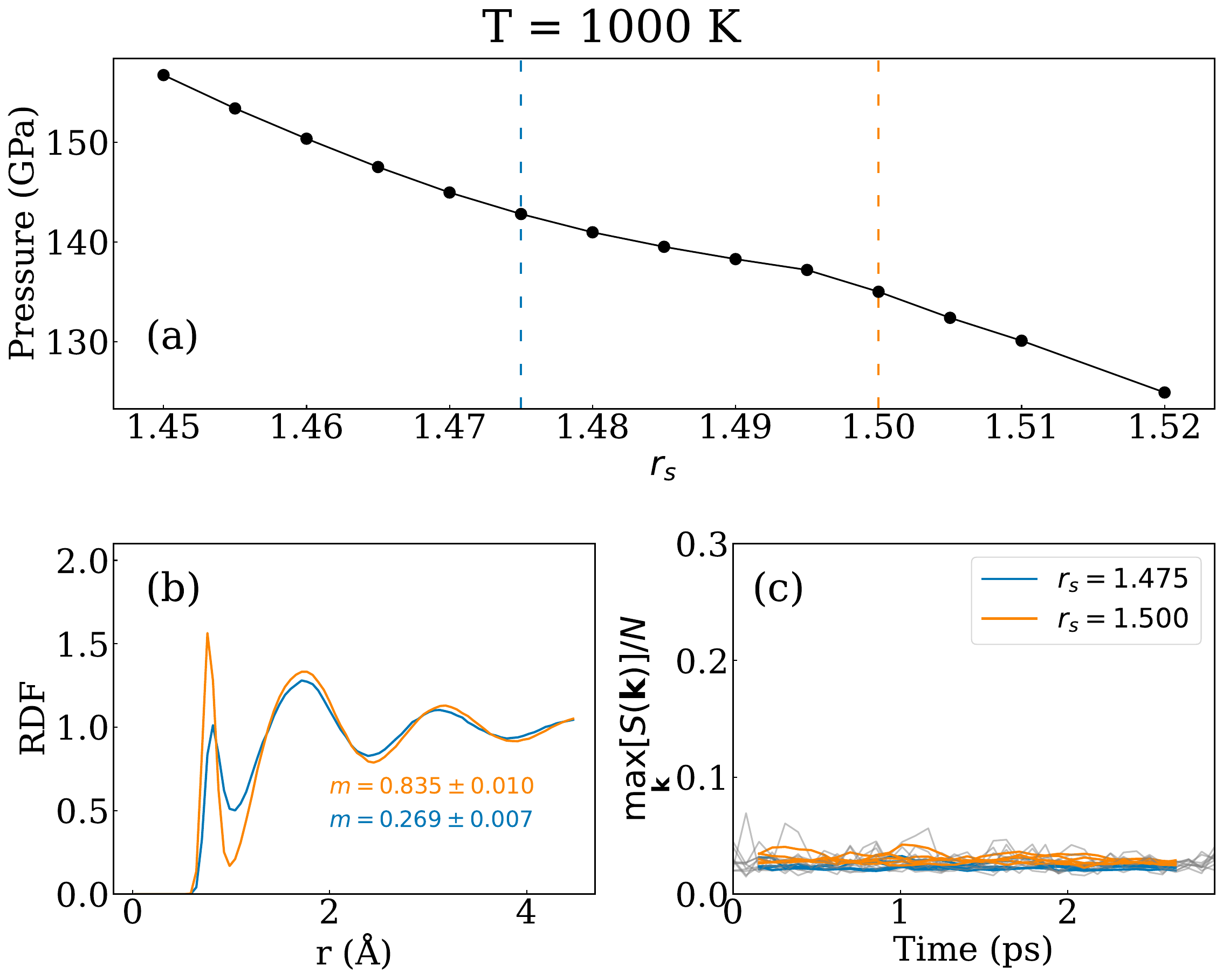}
    \caption{Results of PIOUD simulations on a $512$ hydrogen atoms system at $T = 1000$\,K. (a) Equation of state close to the LLT. The vertical dashed lines highlight the $r_s$ values for which we show additional analysis. (b) Radial distribution function for the two $r_s$ values, i.e.,  $r_s = 1.475$ (blue) and $r_s = 1.50$ (orange). The value of the molecular fraction $m$ (as defined in the main text) is also shown. (c) Value of $\max_{\mathbf{k}} S(\mathbf{k}) /N $ along the trajectories for the two $r_s$ values and its running average on a time window of $375$\,fs. The different lines with the same color correspond to independent runs starting with different initial conditions. }
    \label{fig: PIMD T = 1000K}
\end{figure}

Due to hydrogen light mass, nuclear quantum effects (NQEs) are relevant up to $\sim 3000$~\,K and are known to shift the LLT towards lower pressures~\cite{Pierleoni2016}. Here we investigate whether the physical picture obtained from classical MD changes with the inclusion of NQEs. To do this, we performed constant temperature path integral Ornstein-Uhlenbeck dynamics (PIOUD)~\cite{Mouhat2017} using the MACE model as an energy and force driver. We ran simulations of a system of $N=512$ hydrogen atoms, which is sufficiently large to exclude finite-size effects as shown in the main text. For the dynamics, we used $12$ replicas (beads), a time step of $0.15$~\,fs, and a friction term (used for controlling the temperature) equal to $0.06$~\,fs$^{-1}$. Given the larger computational time needed to perform the PIOUD, to accumulate more data we ran $5$ independent runs for each density and temperature, each starting from different decorrelated initial configurations extracted from the classical trajectories. The equations of state obtained at $T=900$\,K and $T=1000$\,K are shown in Figs.~\ref{fig: PIMD T = 900K} and \ref{fig: PIMD T = 1000K}, respectively. We also calculated the radial distribution function, the molecular fraction, and the structure factor maximum $\max_{\mathbf{k}} S(\mathbf{k})$, for values of the Wigner-Seitz radius $r_s$ before and after the LLT. Although with a different transition pressure, the PIOUD simulations confirm the presence of a first-order phase transition between a molecular crystal and an atomic liquid at $T=900$\,K, followed by a liquid-liquid crossover at larger temperatures (i.e., $T= 1000$\,K).  

\section{Comparison of the MACE model with AIMD}
\begin{figure}[h!]
    \centering
    \includegraphics[width=0.9\linewidth]{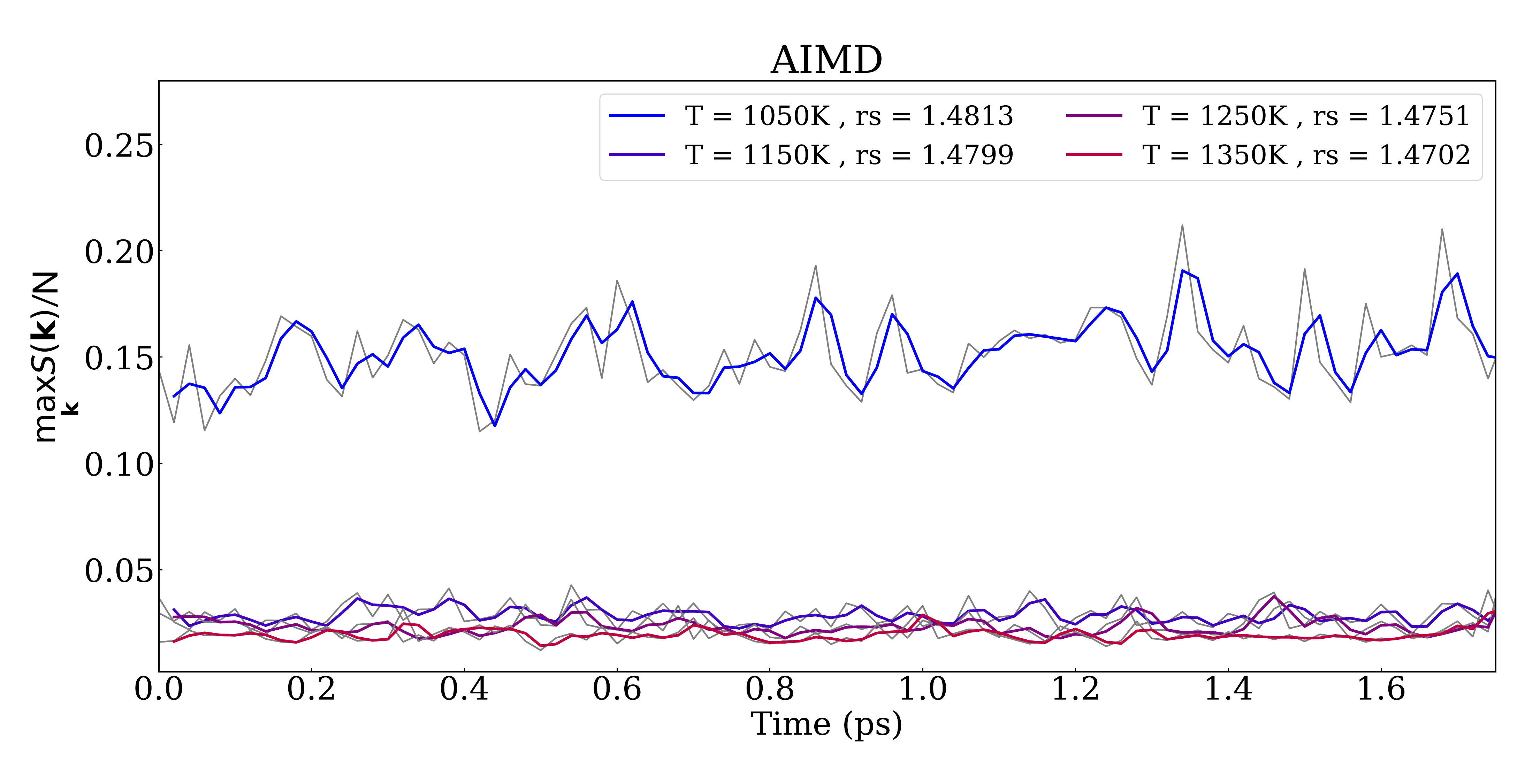}
    \caption{$\max_{\mathbf{k}} S(\mathbf{k})/N$ resolved in time for different values of $r_s$ and temperatures, obtained from AIMD simulations for a system of $512$ atoms. The colored lines are the running averages over $40$\,fs. }
    \label{fig: sk_story_p150 aimd}
\end{figure}

\begin{figure}[h!]
    \centering
    \includegraphics[width=0.9\linewidth]{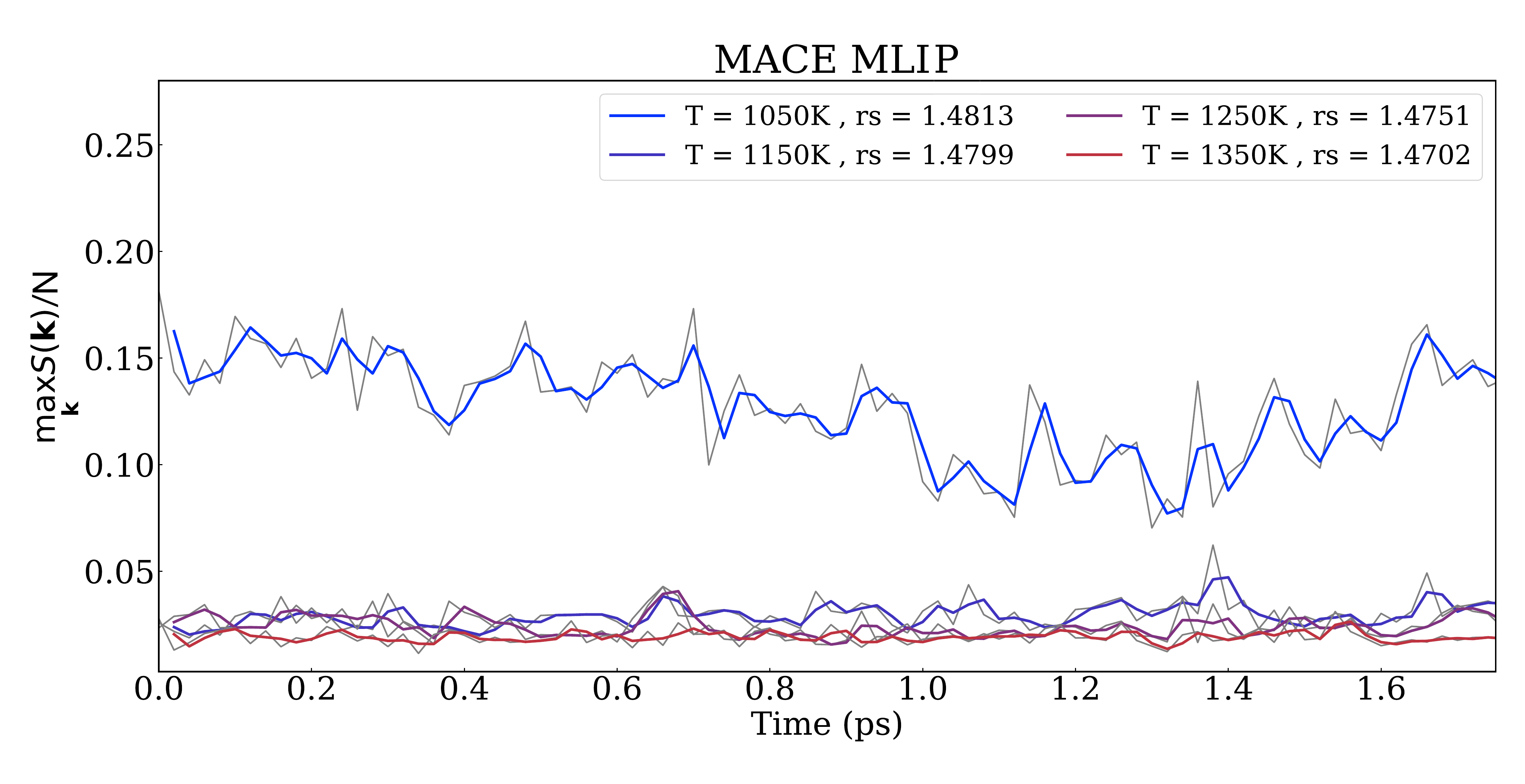}
    \caption{$\max_{\mathbf{k}} S(\mathbf{k}) /N$ resolved in time for different values of $r_s$ and temperatures, obtained from MD simulations using the MACE model for a system of $512$ atoms. The colored lines are the running averages over $40$\,fs.}
    \label{fig: sk_story_p150 mace}
\end{figure}

\begin{figure}[h!]
    \centering
    \includegraphics[width=0.85\linewidth]{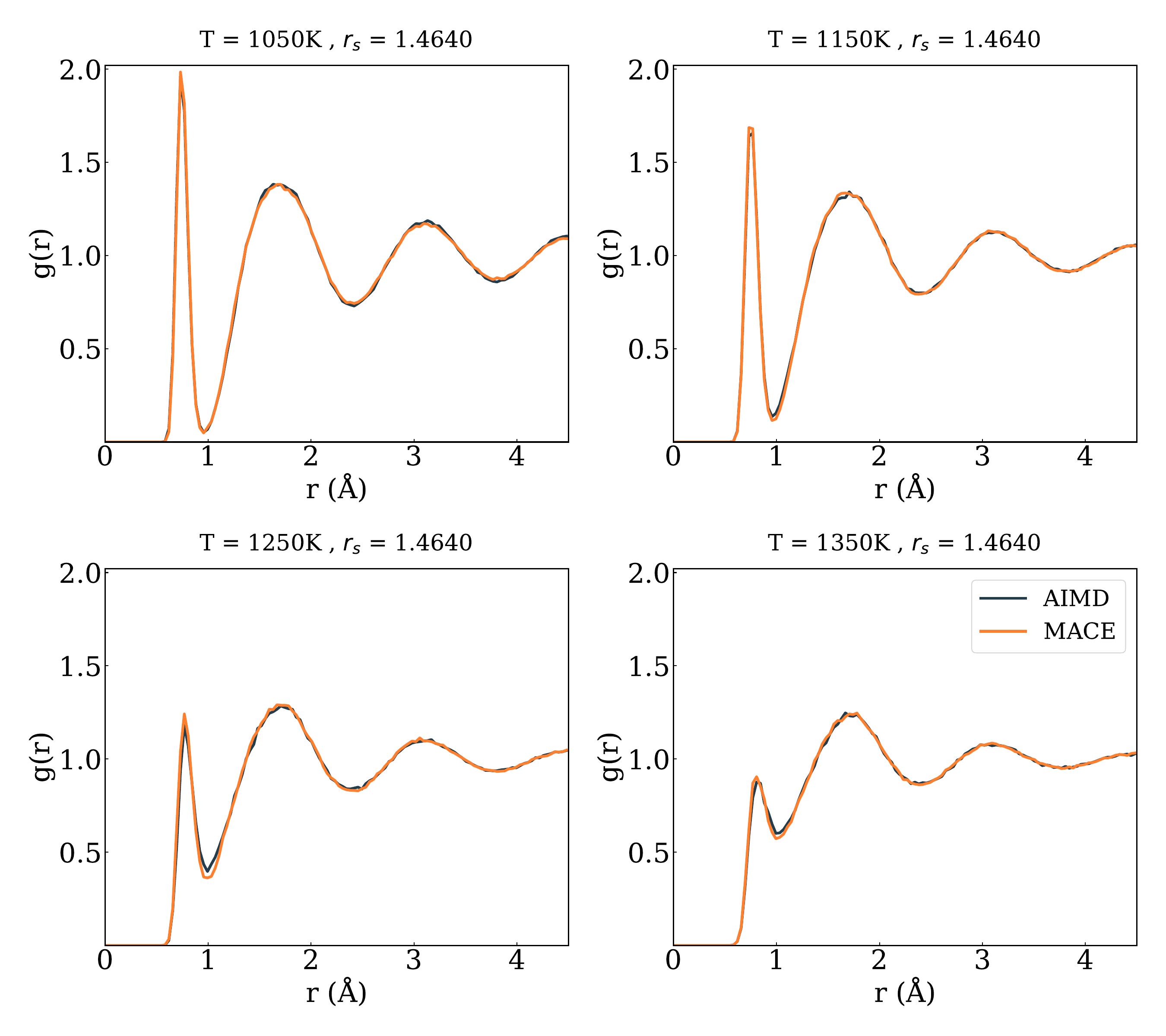}
    \caption{Comparison between AIMD and MACE for the radial distribution function $g(r)$ at different values of $r_s$ and temperatures.}
    \label{fig: rdf comparison AIMD MACE}
\end{figure}

\begin{figure}[h!]
    \centering
    \includegraphics[width=0.7\linewidth]{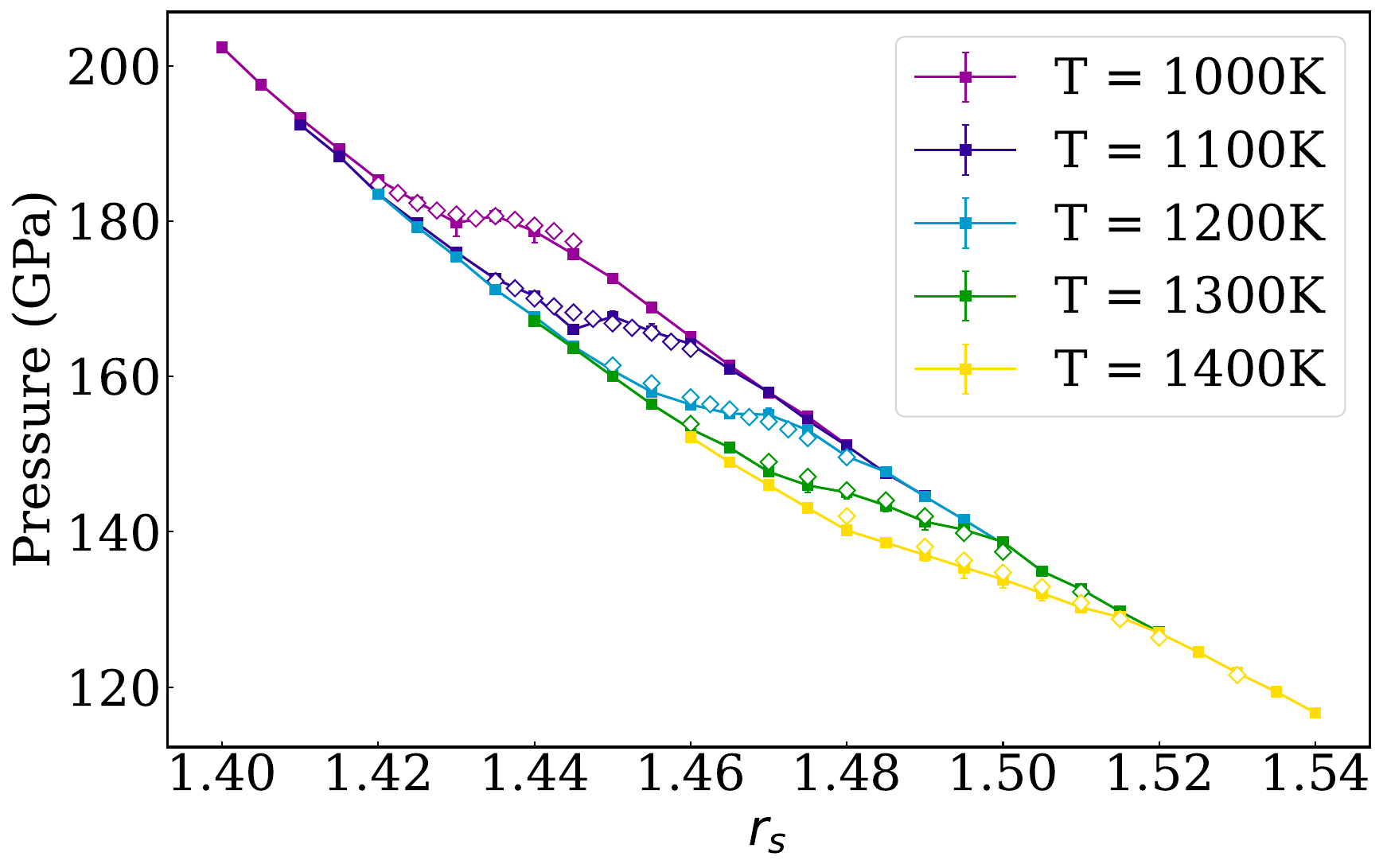}
    \caption{Comparison between the equations of state computed at different temperatures with AIMD (full symbols) and with the final MACE MLIP (empty symbols) for a system of $128$ hydrogen atoms. The AIMD results are taken from \cite{Bischoff2024}.}
    \label{fig: EOS comparison}
\end{figure}

To validate our MLIP, we directly compared the results obtained with the MACE model and AIMD simulations.
In Figs.~(\ref{fig: sk_story_p150 aimd}-\ref{fig: rdf comparison AIMD MACE}), we report NVT results obtained at values of density and temperature roughly corresponding to $p = 150$~GPa taken from Ref.~\cite{Karasiev2021}. We performed simulations of length $\sim 2$~ps with both AIMD and the MACE model on a system of $512$ atoms, and compared the structural properties of the system. 
The time-resolved value of $\max_{\mathbf{k}} S(\mathbf{k})$ for each density and temperature is shown in Fig.~\ref{fig: sk_story_p150 aimd} for AIMD and Fig.~\ref{fig: sk_story_p150 mace} for MACE. 
The two methods show good agreement and both predict solid-like structures at $T=1050$~K. Notice how this temperature matches the value of the melting line of Ref.~\cite{Niu2023} at $p=150$~GPa. 
A comparison of the radial distribution function is reported in Fig.~\ref{fig: rdf comparison AIMD MACE}. At all temperatures, the MACE model shows a remarkable agreement with the AIMD result, both in the molecular and atomic phases. 
Finally, in Fig.~\ref{fig: EOS comparison} we show the equations of state at different values of temperature obtained with the final MACE MLIP and the AIMD ones obtained in Ref.~\cite{Bischoff2024}. Both results considered a system of $128$ hydrogen atoms. The two results are in excellent agreement, with a residual discrepancy which could be due to the short length of the AIMD simulations.

These results further confirm the reliability of our MLIP.

\clearpage

\bibliography{Bibliography}